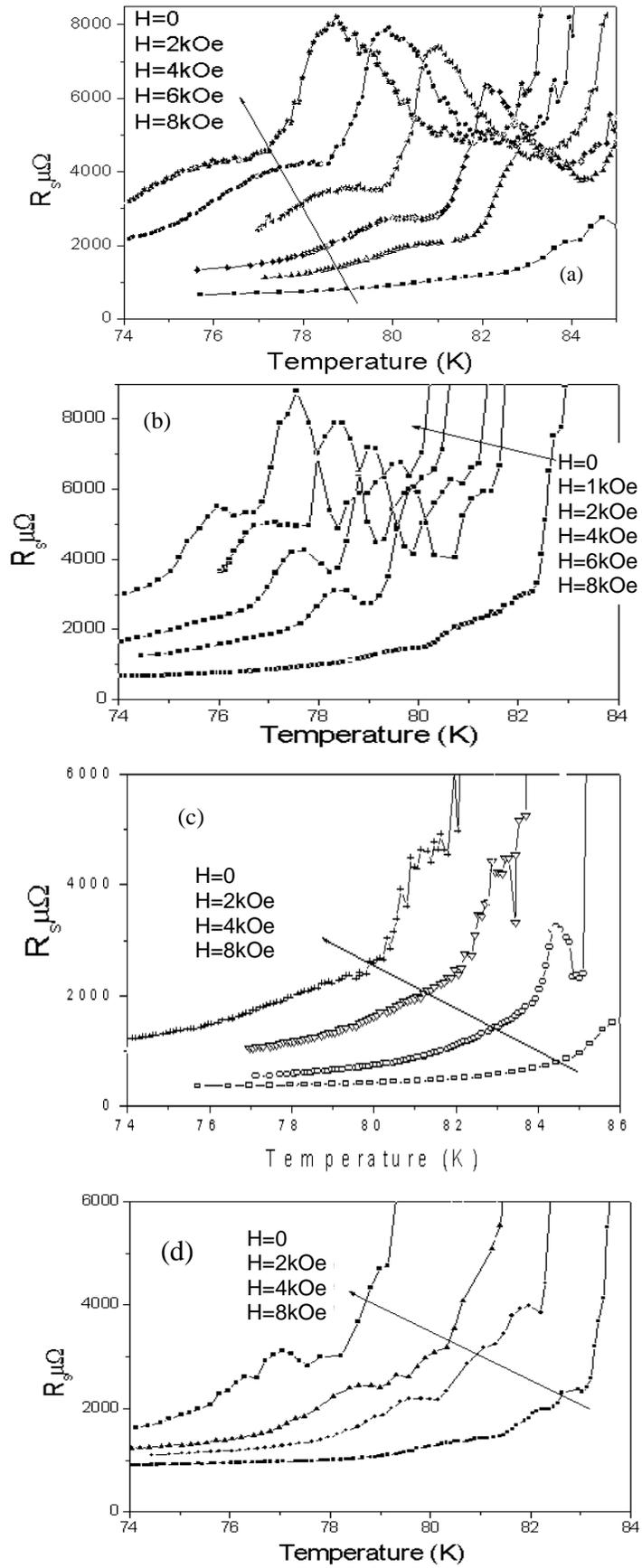

Figure 1 (A. R. Bhangale et al.)

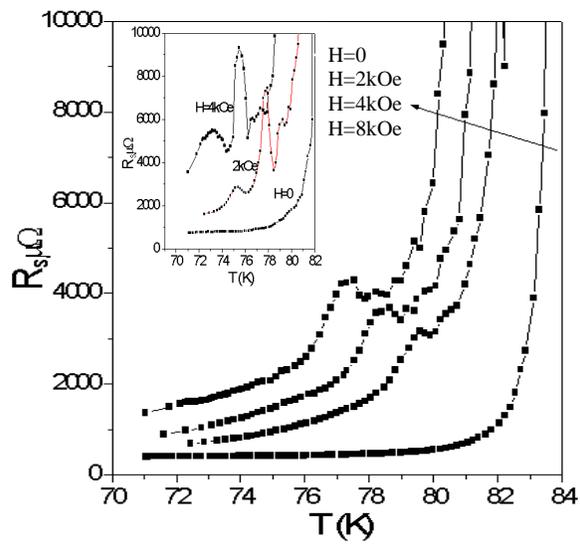

Figure 2 (A. R. Bhangale et al.)

# Peak effect in laser ablated DyBa$_2$Cu$_3$O$_{7-\delta}$ films at microwave frequencies at subcritical currents

A. R. Bhangale[c)], P. Raychaudhuri[a,b)], T. Banerjee[a)], R. Pinto[a)], V. S. Shirodhkar[c)]

[a]Tata Institute of Fundamental Research, Homi Bhabha Rd., Colaba, Mumbai 400005, India.
[b]School of Physics and Astronomy, University of Birmingham, Edgbaston, Birmingham B15 2TT, England.
[c]Department of Physics, Institute of Science, 15 Madam Cama Rd., Mumbai 400032, India.

*Abstract:* In this paper we report the observation of a peak in the microwave surface resistance (at frequencies ~10GHz) of laser ablated DyBa$_2$Cu$_3$O$_{7-\delta}$ films in magnetic field ranging from 2 to 9kOe ( $\|c$ ) close to the superconducting transition temperature (T$_c$(H)). The exact nature of peak is sample dependent but it follows a general behaviour. The peak shifts to lower temperature when the magnetic field is increased. It has strong frequency dependence and the peak is pronounced at frequencies close to the depinning frequency of the flux line lattice. From the observed temperature and field dependence we argue that this peak is associated with the order disorder transition of the flux line lattice close to T$_c$(H).

The phenomenon of 'peak effect' associated with a peak in the critical current ($J_c$) below the superconductor-normal boundary in type-II superconductors has attracted widespread attention in order to understand the order-disorder transition[1] in the flux line lattice (FLL) in type-II superconductors in the presence of pinning arising from defects in crystallographic lattice. In varying temperature or field measurement this peak is manifested in both transport and magnetic measurements such as resistivity[1], magnetization[2] or a.c. susceptibility[3]. Whereas the static measurements such as magnetization probe the response of the static FLL to an external applied field or temperature, the dynamics of the vortex lattice is probed by applied a small time dependent perturbation and measuring the response of the FLL or by measuring the transport properties above the critical current ($J_c(H)$) when the flux lines are moving. Both kinds of measurements have yielded useful information in recent times regarding the process of disordering of the FLL.

The peak in the critical currents occurring due to the order-disorder transition of the FLL is rationalized within the Larkin-Ovchinikov[4] scenario where the effective pinning force on the FLL is given by $BJ_c(H)=(n_p<f^2>/V_c)^{1/2}$ where $n_p$ is the density of pinning centers, $f$ is the elementary pinning force parameter, $V_c$ is the Larkin volume over which the FLL maintains its spatial order. With increasing temperature or field $<f^2>$ decreases thereby decreasing the critical current density. However, at an order-disorder transition of the FLL $V_c$ goes to zero thereby giving rise to peak in $J_c(H)$. The exact details of the order-disorder transition has remained an issue of intense debate. Several scenario, such as the melting of the FLL[5] or its amorphization to a state with quenched random disorder[6] have been proposed to as the mechanism governing the peak effect phenomenon in type II superconductors.

In typical ac susceptibility measurements used to study the dynamics of the FLL the measurement frequencies vary from a few Hertz to few MHz. These measurements do not reveal any frequency dependence of the peak effect in agreement with a true phase transition[7]. There are however no reports of peak effect in microwave frequencies. Recently we have been able to observe a pronounced peak effect in epitaxial thin films of $DyBa_2Cu_3O_{7-\delta}$ (DBCO) at gigaHertz frequencies[8]. In a typical microwave measurement the small microwave excitation induces a current which is smaller than the critical current of the superconductor. The vortices therefore move back and forth close to the minimum of the pinning potential (V) and experiences a restoring force close to the potential minimum. The motion at these frequencies is described by the equation of motion of a massless harmonic oscillator in a viscous medium given by[9],

$$\eta \dot{x} + kx = F \quad (1)$$

where $\eta$ is the Bardeen-Stephen viscous drag coefficient, $k$ is the pinning force constant, and F is the external force on the vortex which is given by $F=J\Phi_0$. It can be shown that the vortex impedance is given by,

$$Z_v = \frac{\Phi_0 H}{\eta} \frac{1}{(1+i\omega_p/\omega)} \quad (2)$$

where $\omega_p$, is the depinning frequency given by, $\omega_p=(k/\eta)$. It follows from equation (2) that the vortex motion crosses over from inductive for $\omega \ll \omega_p$ to resistive for frequencies $\omega \gg \omega_p$ giving rise to dissipation. To understand the origin of the peak using equation (2) we have to note that in the collective pinning scenario the vortices within a Larkin volume $V_c$ move together like a semi-rigid body. Therefore the effective restoring force constant on the vortices will be given by the effective restoring force on all the vortices in $V_c$. Therefore the effective restoring force constant on the vortices within $V_c$ will be given by, $k \propto (n_p <f^2>/V_c)^{1/2}$ and will show a peak at order disorder transition of the FLL[9]. Since $\eta$ decreases monotonically with increase in temperature, this will give rise to a peak in $\omega_p$. Therefore if the measurement frequency $\omega$ is close to the depinning frequency the surface resistance will show a dip (corresponding to the peak in $\omega_p$) as one crosses the order-disorder transition. On the other hand this effect will be much less prominent if the measurements are done much below or above the depinning frequency, namely, for $\omega >> \omega_p$ or $\omega << \omega_p$. In high $T_c$ cuprates with 123 structure the depinning frequency has been reported to vary in the range 5-40GHz[10].

In this paper we report the microwave response of two superconducting $DyBa_2Cu_3O_{7-\delta}$ films (with *c* perpendicular to substrate) grown by pulsed laser deposition on single crystalline $LaAlO_3$ substrates. They are denoted as samples S1 and S2. The films were grown in 250mTorr ambient oxygen pressure. The substrate temperature was kept between 750-780$^0$C and the target substrate temperature was fixed at 4cm. The $T_c$ of the both films measured through ac susceptibility using flat coil geometry was 90±0.2K. The films were subsequently patterned into striplines with a width of 150μm and length 10mm. The measurements of microwave surface resistance were carried out using stripline resonator technique using a Helwett-Packard scalar network analyzer[11]. All the meaurements were carried out at the same relative incident microwave power level.

Magnetic fields up to 9kOe was applied parallel to the *c* axis using a conventional electromagnet.

Figure 1(a)-(d) show the surface resistance ($R_s$) as a function of temperature measured at 4.75GHz and 9.5GHz (corresponding to the fundamental and 1$^{st}$ harmonic excitation of the striplines) of the S1 and S2 respectively. Both display a pronounced peak followed by a dip in the $R_s$ measured at 9.5GHz. The evolution of the peak in $R_s$ with magnetic field show a common behaviour in both the samples. With increasing magnetic field $R_s$ increases and the peak shifts to lower temperatures. For the $R_s$ measured at 4.75GHz the peak becomes much less prominent. However, the peak in $R_s$ at 9.5GHz is sharper in the sample S2 than in S1. On the other hand the peak in $R_s$ at 4.75 GHz is more pronounced in S1 than S2. Also in addition to the main peak, the surface resistance at 9.5GHz shows a broad hump like structure before the main peak. This kind of complex structure is not uncommon near the peak effect region and has been reported earlier in ac susceptibility and transport measurements.

To understand the difference between S1 and S2 we first note that though samples S1 and S2 have nominally the same $T_c$ they contain different amount of disorder since they were grown in separate runs. Consequently, $n_p$ and therefore the effective $k$ and $\omega_p$ are different in the two samples. We have shown earlier that the peak in microwave will be most prominent close to $\omega_p$. From the present results it can be argued that $\omega_p$ is smaller in S1 than S2 though both lie in the range of few GHz to tens of GHz since at lower frequencies the peak is more prominent in S1 and at higher frequencies it is more prominent is S2. This implies that S1 has a smaller density of disorder than S2 which is seen from the value of the $R_s$ at 75K at 4.75GHz in zero field which is 300µΩ for S1 and 900µΩ for S2. It is however to be noted that equation (2) is not strictly valid for temperatures close to $T_c$ where large scale motion of the vortices can take place due to thermal excitation. This could be in part the reason why the peaks at 4.75GHz and 9.5GHz do not occur at the same temperature and fields.

In order to check whether the magnetic ion Dy has any particular role in the peak effect observed in microwaves we have also carried out these measurements in a laser ablated film of the isostructural superconductor $YBa_2Cu_3O_{7-\delta}$ (YBCO) (Figure 2). The peak effect in the YBCO film is similar to the peak effect in S2. The peak in the $R_s$ at 9.5GHz is sharp. Though less pronounced a clear peak effect is observed in the $R_s$ at 4.75GHz.

In summary we have observed a pronounced peak effect in the microwave surface resistance of superconducting 123 films in the presence of a dc magnetic field. From the temperature and field dependence of the peak we have argued that the peak is due to the order-disorder transition of the vortex lattice. Though related, this however is different from the peak in $J_c$ observed from transport or magnetization measurements since the microwave excitation induces a current which is smaller than the critical current of the superconductor. From our observation we also conclude that the depinning frequency is ~10GHz which matches reasonably well with earlier estimates of depinning frequency in YBCO[10].

**Figure Captions:**

1. $R_s$ versus temperature at 9.5GHz for (a) S1 and (b) S2 for different fields. Same at 4.75GHz for (c) S1 and (d) S2.

2. $R_s$ versus temperature for the YBCO film at 4.75GHz at different fields; the *inset* shows the same for 9.5GHz.